\documentstyle[aps,graphicx]{revtex}

\def\r{{\bf r}}
\def\px{{\partial \over \partial x}}
\def\pt{{\partial \over \partial t}}
\def\bra{\langle}  \def\ket{\rangle}
\def\hpn{\hat \psi} \def\hpd{\hat \psi^\dagger}
\def\efn{\hat \psi (\r)} \def\efd{\hat \psi^\dagger (\r)}
 
\def\rfn{\hat \psi_{\rm R} (\r)} 
\def\ofn{\hat \psi_1 (x)} \def\ofd{\hat \psi_1^\dagger (x)}
\def\px{{\partial \over \partial x}}
\def\pt{{\partial \over \partial t}}

\begin{document}

\title{\bf
Nonequilibrium Mesoscopic Conductors
Driven by Reservoirs
}
\author{
Akira Shimizu 
and Hiroaki Kato 
\\
{\small Department of Basic Science, University of Tokyo, 
Komaba, Tokyo 153-8902, Japan}
}

\date{}

\maketitle              

%
\vspace{-60mm}
\begin{center}
{\small
Proceedings of 219th International WEH Workshop in Hamburg, Germany, 1999

{\it
Interactions and Quantum Transport Properties of Lower Dimensional Systems
} 

(ed. T. Brandes, Springer-Verlag, 2000)
}
\end{center}
\vspace{35mm}

\begin{abstract}
In order to specify a nonequilibrium steady state
of a quantum wire (QWR), one must connect reservoirs to it.
Since reservoirs should be large 2d or 3d systems,
the total system is a large and inhomogeneous 2d or 3d system,
in which $e$-$e$ interactions have the same strength in 
all regions.
However, most theories of interacting electrons in QWR
considered simplified 1d models, in 
which reservoirs are absent or replaced with noninteracting 
1d leads.
We first discuss fundamental problems of such theories
in view of nonequilibrium statistical mechanics.
We then present formulations which are free from such difficulties, 
and discuss what is going on in 
mesoscopic systems in nonequilibrium steady state.
In particular, we point out important roles of 
energy corrections and non-mechanical forces, which 
are induced by a finite current.
\end{abstract}

\section{Introduction}
\label{sec_intro}

\begin{figure}[t]
\begin{center}
\includegraphics[width=.9\textwidth]{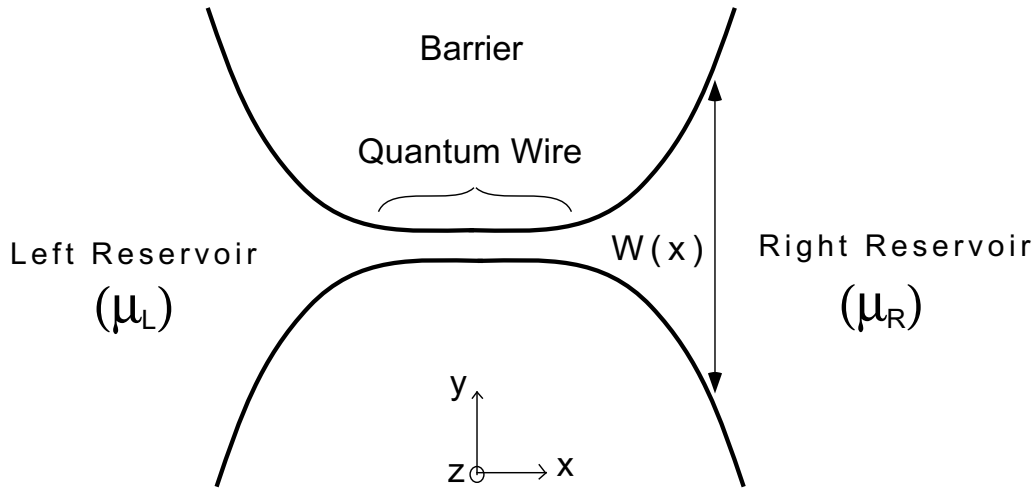}
\end{center}
\caption[]{A two-terminal conductor 
composed of a QWR and reservoirs.
}
\label{fig1}
\end{figure}

According to nonequilibrium thermodynamics, 
one can specify 
nonequilibrium states of macroscopic systems 
by specifying local values of thermodynamical 
quantities, such as the local density and the local 
temperature, 
because of the local equilibrium \cite{callen,zubarev}.
When one studies 
transport properties of 
a mesoscopic conductor (quantum wire (QWR)), however, 
the local equilibrium is not realized in it, 
because it is too small.
Hence, in order to specify its nonequilibrium state uniquely,
one must connect {\em reservoirs} to it,
and specify their chemical potentials ($\mu_{\rm L}$, $\mu_{\rm R}$)
instead of
specifying the local quantities of the conductor (Fig.\ \ref{fig1}).
The reservoirs should be large (macroscopic) 2d or 3d systems.
Therefore, 
to really understand transport properties, 
we must analyze such a composite system of 
the QWR and the 2d or 3d reservoirs,
Although the QWR itself 
may be a homogeneous 1d system,  
the total system is a {\em 2d or 3d inhomogeneous system} 
without the translational symmetry.
Moreover, {\em many-body interactions} are important 
{\em both} in the conductor and in the reservoirs:
If electrons were free in a reservoir, neither could electrons   
be injected (absorbed) into (from) the conductor, nor could they 
relax to achieve the local equilibrium.
However, 
most theories considered simplified 1d models, in 
which reservoirs are absent or replaced with noninteracting 
1d leads 
\cite{AR,KF,FN,OF,su92,maslov,pono,kawabata,safi,oreg96}.

In this paper, we study  transport properties of 
a composite system of a QWR plus reservoirs, 
where $e$-$e$ interactions are present in all regions.
By critically reviewing theories of the conductance, 
we first point out fundamental problems of the theories in view 
of nonequilibrium statistical mechanics.
We then present formulations which are free from such difficulties, 
and discuss what is going on in mesoscopic systems in nonequilibrium steady 
state.
In particular, we point out important roles of 
energy corrections and non-mechanical forces, which 
are induced by a finite current.

\section{A critical review of theories of the DC conductance}
\label{sec_review}

In this section, we critically review theories of 
the DC conductance $G$ of interacting electrons in a QWR.
Note that two theories which predict different nonequilibrium states
can (be adjusted to) give the same value of $G$ (to agree with experiment).
Hence, the comparison of the values of $G$ among different theories 
is not sufficient.
%
For definiteness, 
we consider a two-terminal conductor 
composed of a quantum wire (QWR) and two reservoirs 
(Fig.\ \ref{fig1}), which are defined by a 
confining potential $u^{\rm c}$,
at zero temperature.
Throughout this paper, {\it we assume that 
$u^{\rm c}$ is smooth and slowly-varying, 
so that electrons are not reflected 
by $u^{\rm c}$} (i.e., the wavefunction evolves adiabatically).
We also assume that only the lowest subband of the QWR is 
occupied by electrons.
A finite current $I$ is induced by applying 
a finite difference 
$\Delta \mu = \mu_{\rm L} - \mu_{\rm R}$ of chemical potentials 
between the two reservoirs, 
and the DC conductance is defined by 
$
G \equiv \langle I \rangle/(\Delta \mu/e)
$ \cite{why_e},
where $\langle I \rangle$ is the average value of $I$. 

Let us consider a clean QWR, which has no impurities or defects.
For {\it non-interacting} electrons 
the Landauer-B\"uttiker formula
gives $G= e^2/ \pi \hbar$ \cite{LB}, whereas
$G$ for {\it interacting} electrons
has been a subject of controversy \cite{nontrivial}.
%
Most theories before 1995 \cite{AR,KF,FN,OF}
predicted that $G$ should be 
``renormalized" by
the $e$-$e$ interactions as
$G= K_\rho e^2/ \pi \hbar$,
where $K_\rho$ is a parameter characterizing 
the Tomonaga-Luttinger liquid (TLL) 
\cite{tomonaga,luttinger,haldane,KY}.
However, Tarucha et al.\ found experimentally that
$G \simeq e^2/ \pi \hbar$ for a QWR of $K_\rho \simeq 0.7$ \cite{tarucha}.
Then, several theoretical papers have been published 
to explain the absence of the renormalization of $G$
\cite{maslov,pono,kawabata,safi,oreg96,s96}.
Although they concluded the same result, $G= e^2/ \pi \hbar$, 
the theoretical frameworks and the physics are 
very different from each other.
Since most theories are based either on the Kubo formula \cite{kubo}
(or, similar ones based on the adiabatic switching of an 
``external" field), 
or on the scattering theory, we review these two types of theories 
critically in this section.

\subsection{
Problems and limitations of the Kubo 
formula when it is 
applied to mesoscopic conductors
}
\label{sec_pl}

When one considers a physical system, it always 
interacts with other systems, R$_1$, R$_2$, $\cdots$,  
which are called heat baths or reservoirs.
Nonequilibrium properties of the system can be calculated 
if one knows the reduced density matrix 
$\hat \zeta \equiv {\rm Tr}_{\rm R1+R2+\cdots} 
[ \hat \zeta_{\rm total} ]$.
Here,
$\hat \zeta_{\rm total}$ is the density operator of 
the total system, and 
${\rm Tr}_{\rm R1+R2+\cdots}$ denotes the trace operation over reservoirs'
degrees of freedom.
To find $\hat \zeta$, Kubo \cite{kubo} 
assumed that the system is initially in 
its equilibrium state.
Then an ``external field'' ${\bf E}_{\rm ext}$ is applied adiabatically
(i.e., ${\bf E}_{\rm ext} \propto e^{-\epsilon |t|}$),
{\em which is a fictitious field} because 
it does not always have its physical correspondence
(see below).
The time evolution of $\hat \zeta$ was calculated using
the von Neumann equation of an isolated system; i.e., 
{\em it was assumed that the system were isolated from the reservoirs} 
during the time evolution \cite{zubarev}.
Because of these two assumptions (the fictitious field and 
isolated system), 
some conditions are required to get correct results 
by the Kubo formula.
To examine the conditions, 
we must distinguish between 
{\em non-dissipative responses} (such as the DC magnetic susceptibility)
and 
{\em dissipative responses} (such as the DC conductivity $\sigma$).
The non-dissipative responses are essentially 
equilibrium properties of the system; in fact, 
they can be calculated from {\it equilibrium} statistical mechanics.

For non-dissipative responses,
Kubo \cite{kubo,KTH} and Suzuki \cite{suzuki}
established the conditions for the validity of 
the Kubo formula,
by comparing the formula with the results of 
equilibrium statistical mechanics:
(i) The proper order should be taken in the limiting procedures
of $\omega, q \to 0$ and $V \to \infty$, 
where $\omega$ and $q$ are the frequency and wavenumber of 
the external field, and $V$ denotes the system volume.
(ii) The dynamics of the system should have the following 
property;
\begin{equation}
\lim_{t \to \infty}
\langle 
\hat A \hat B(t)
\rangle_{\rm eq}
=
\langle \hat A \rangle_{\rm eq} \langle \hat B \rangle_{\rm eq},
\label{mixing_eq}\end{equation}
where $\langle \cdots \rangle_{\rm eq}$
denotes the expectation value in the thermal equilibrium, 
and $\hat A$ and $\hat B$ are the operators whose 
correlation is evaluated in the Kubo formula.
Any integrable models do not have 
this property \cite{suzuki,ott,nakano,dset}.
Hence, {\it the Kubo formula is not applicable 
to integrable models, such as the Luttinger model}, 
even for (the simple case of) non-dissipative responses \cite{suzuki}.

%

\begin{figure}[t]
\begin{center}
\includegraphics[width=.7\textwidth]{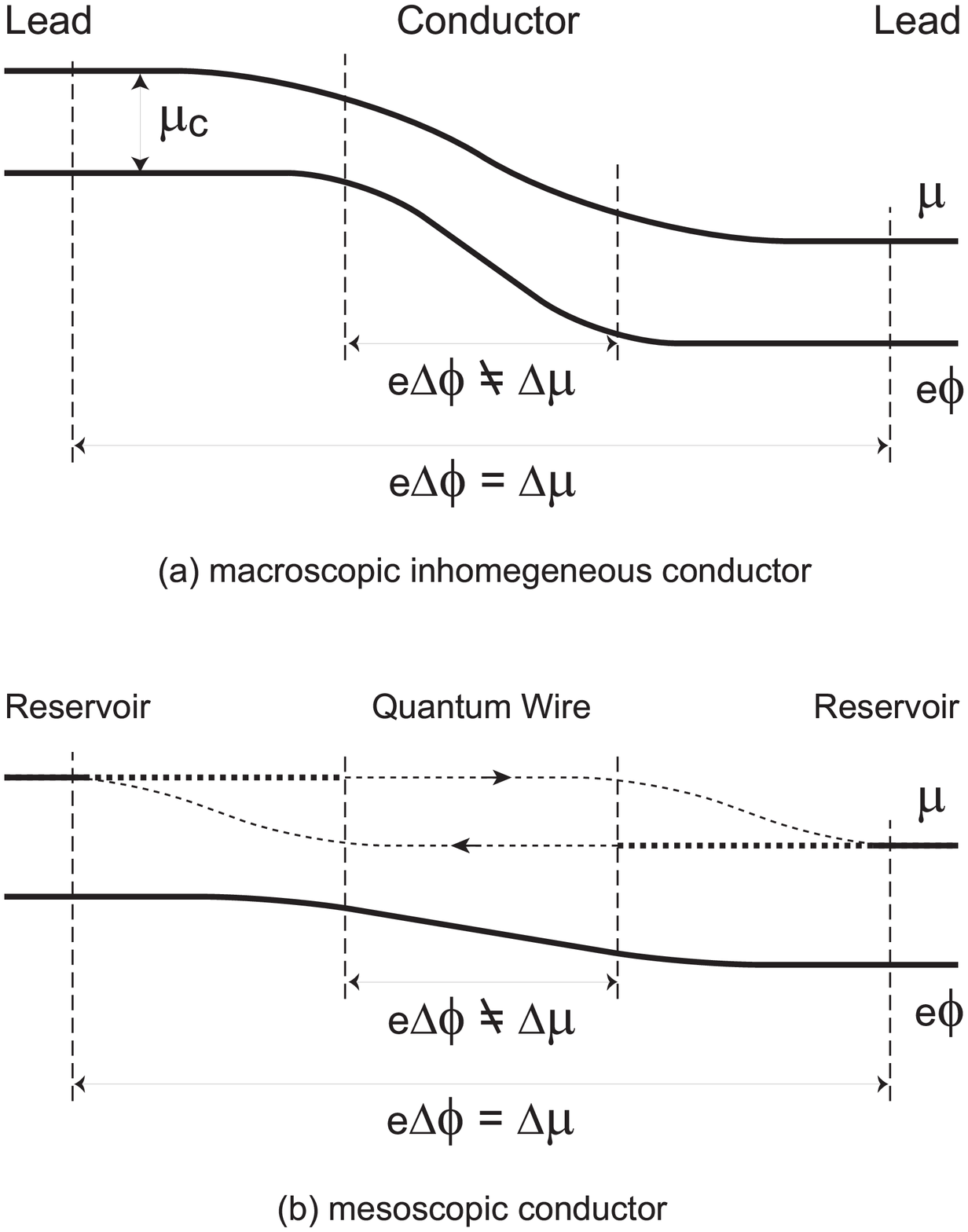}
\end{center}
\caption[]{Schematic plots of the chemical potential 
$\mu$ \cite{def_of_mu} and the electrostatic potential $\phi$, for
(a) a macroscopic inhomogeneous conductor and
(b) a mesoscopic conductor.
For case (a), the 
local equilibrium is established, and thus 
$\mu$ and $\phi$ can be defined in all regions. 
The differences $e \Delta \phi$ and $\Delta \mu$ 
are not equal if one takes the differences between 
both ends of the conductor, 
whereas $e \Delta \phi = \Delta \mu$ 
if the differences are taken between 
the leads. 
For case (b), $\mu$ cannot be defined 
in the QWR and boundary regions 
(although in some cases $\mu$ could be defined separately 
for left- and right-going electrons), 
whereas
$\phi$ can be defined in all regions.
Similarly to case (a), $e \Delta \phi \neq \Delta \mu$ 
if one takes the differences between 
both ends of the QWR, 
whereas $e \Delta \phi = \Delta \mu$ 
if the differences are taken between 
the reservoirs.
}
\label{fig_inhomo}
\end{figure}

For dissipative responses, 
the conditions for the applicability of the Kubo formula
would be stronger.
Unfortunately, however, 
they are not completely clarified, and we here list 
some of known or suggested conditions for $\sigma$:

\noindent
(i') Like as condition (i), 
the proper order should be taken in the limiting procedures.
For $\sigma$ the order should be \cite{mahan}
\begin{equation}
\sigma =
\lim_{\omega \to 0} \lim_{q \to 0} \lim_{V \to \infty}
\sigma_{\rm formula} (q, \omega; V).
\end{equation}
\noindent
(ii')  Concerning condition (ii), 
a stronger condition seems necessary for 
dissipative responses:
The closed system that is taken in the calculation of
the Kubo formula should have the thermodynamical
stability, i.e., 
it approaches the thermal equilibrium when 
it is initially subject to a macroscopic perturbation.
(Otherwise, it would be unlikely for the system 
to approach the correct steady state in the presence of 
an external field.)
In classical Hamiltonian systems, this condition is 
almost equivalent to 
the ``mixing property'' \cite{ott,nakano,dset}, 
which states that 
Eq.\ (\ref{mixing_eq}) should hold 
for {\it any} $\hat A$ and $\hat B$, 
where $\langle \cdots \rangle_{\rm eq}$
is now taken as the average over the equi-energy surface.
It is this condition, rather than the ``ergodicity'', that  
guarantees the thermodynamical stability \cite{ott,nakano,dset}.
Although real physical systems should always have 
this property, some theoretical models do not.
In particular, 
any integrable models do not have 
this property \cite{ott,nakano,dset}. 

\noindent
(iii') We here suggest that all driving forces, 
including non-mechanical ones, should be identified \cite{shmz_unpub}.
In fact, 
the formula gives the current density in the following form,
\begin{equation}
\langle
{\bf J} 
\rangle
=
\sigma_{\rm formula} {\bf E}_{\rm ext},
\end{equation}
whereas the {\it exact} definition of $\sigma$ is given by 
nonequilibrium thermodynamics as \cite{callen,zubarev}
\begin{equation}
\langle
{\bf J} 
\rangle
=
- \sigma \nabla (\mu/e)- L_{12} \nabla \beta
=
\sigma {\bf E} - \sigma \nabla (\mu_{\rm c}/e)- L_{12} \nabla \beta.
\label{JNETD}\end{equation}
Here, $\beta$ denotes the inverse temperature, 
$\mu$ is the ``chemical potential'' 
which consists of a chemical portion $\mu_{\rm c}$ 
and
the electrostatic potential $\phi$ \cite{callen,def_of_mu};
\begin{equation}
\mu = \mu_{\rm c} + e \phi 
\quad
\mbox{(hence, $\Delta \mu = \Delta \mu_{\rm c} + e \Delta \phi$ 
for differences)}.
\label{muandmuc}\end{equation}
Hence, to evaluate $\sigma$, one must find the relation between 
${\bf E}_{\rm ext}$ and 
${\bf E}$, $\nabla \mu_{\rm c}$ and $\nabla \beta$.
In {\it homogeneous} systems, it is expected that 
$\nabla \mu_{\rm c} = \nabla \beta = 0$, hence 
it is sufficient to find the relation between 
the fictitious field ${\bf E}_{\rm ext}$ and 
the real field ${\bf E}$ \cite{zubarev,kawabata,izuyama}.
In {\it inhomogeneous} systems,  however, 
$\nabla \mu_{\rm c} \neq 0$ and/or $\nabla \beta \neq 0$ in general \cite{pn},
as shown in Fig.\ \ref{fig_inhomo} (a).
Therefore, 
one must find the relation between ${\bf E}_{\rm ext}$
and these ``non-mechanical forces'' \cite{shmz_unpub,nmf}.
(See section \ref{sec_nmf}.)

Unfortunately, these conditions are not satisfied 
in theories based on simplified models of mesoscopic systems.
For example, the Luttinger model \cite{luttinger} used in much literature
does not satisfy conditions (i) and (ii)
because it is integrable.
To get reasonable results, 
subtle procedures, which have not been justified yet, were
taken in actual calculations.
Moreover, the non-mechanical forces have not been examined, although
they would be important because
a mesoscopic conductor (a QWR plus reservoirs) is 
an inhomogeneous system.

We also mention limitation of the Kubo formula:
it cannot be applied to the nonequilibrium noise (NEN), 
which is the current fluctuation in the presence of a finite current
$\langle I \rangle$ ($= G \Delta \mu$)
\cite{su92,isqm92,NENth1,NENth2,NENex1,NENex2}.
The NEN at low frequency, 
$\langle \delta I^2 \rangle^{\omega \simeq 0}$, 
is usually proportional to 
$|\langle I \rangle| \propto | \Delta \mu |$.
However, the Kubo formula 
assumes power series expansion 
about $\Delta \mu = 0$,
hence cannot give any function of $|\Delta \mu|$
\cite{shmz_unpub,exception}.

%

In sections \ref{sec_combined} and \ref{sec_proj}, 
we present other formulations which are free from 
these problems and limitations.
These formulations clarify what is going on in 
nonequilibrium mesoscopic conductors, because 
one can find the nonequilibrium steady state. 
This is impossible by 
the Kubo formula because it evaluates correlation functions 
in the \emph{equilibrium} state.

\subsection{Scattering-theoretical approaches} 
\label{sec_scatt}

In view of many problems and limitations of the Kubo formula,
it is natural to try to generalize Landauer's theory \cite{LB} to 
treat conductors with many-body interactions.
Namely, the DC conductance may be given in terms of
the scattering matrix (S matrix) for interacting electrons
\cite{su92,safi,isqm92}.

The advantages of the scattering-theoretical approaches
may be as follows:
(i) Neither the translation of
$\Delta \phi_{\rm ext}$ into $\Delta \mu$ 
nor the subtle limiting procedures of $\omega, q$ and $V$ 
is necessary.
(ii) There is no need for the mixing property of 
the 1d Hamiltonian $\hat H_1$. Hence, 
$\hat H_1$ can be the Hamiltonian of 
integrable 1d systems such as the TLL.
(iii) In contrast to the Kubo formula, 
one can calculate the NEN \cite{su92,isqm92,NENth1,NENth2}.

However, to define the S matrix, 
one must define incoming and outgoing states.
Although they can be defined trivially for free electrons, 
it is nontrivial in the presence of many-body interactions.
In high-energy physics, 
they are defined based on the {\it asymptotic condition},
which assumes that particles behave like free (but renormalized) ones
as $t \to \pm \infty$, i.e., before and after the collision
\cite{haag}.
For example, an electron (in the vacuum) before or after the collision 
becomes a localized ``cloud" of electrons and positrons, 
which extend only over the Compton length, 
and this cloud can be regarded as a renormalized electron.
In condensed-matter physics, on the other hand, 
the asymptotic condition is not satisfied for 
electrons in metals and doped semiconductors.
In fact, elementary excitations (Landau's quasi particles) are 
accompanied with the backflow,
which extends {\it all over the crystal} \cite{nozieres}, 
in contradiction to the asymptotic condition.
Because of this fundamental difficulty,
the scattering approaches to mesoscopic 
conductors replaced the reservoirs with 1d leads
in which electrons are free \cite{su92,safi,isqm92}.
Therefore, 
real reservoirs, in which electrons behave as 2d or 3d 
interacting electrons, 
have not been treated by the scattering-theoretical approaches.

\section{
Combined Use of Microscopic Theory and Thermodynamics
}\label{sec_combined}
The basic idea of this method \cite{s96}
is as follows:
Since a QWR is a small system, and is most important, 
it should be treated with a full quantum theory.
On the other hand, 
reservoirs are large systems whose dynamics is complicated, 
hence it could be treated with thermodynamics (in a wide sense).
Utilizing these observations, 
we shall develop thermodynamical arguments
to find 
the nonequilibrium steady state that is realized when 
a finite $\Delta \mu$ is applied between 
the reservoirs.
This is the key of this method because
when the steady state is found, $G$ (and other observables) can be 
calculated by straightforward calculations.
Although in some cases {\it formal} calculations can be performed
without finding the steady state \cite{oreg96}, 
we stress that such formal theories are incomplete 
because another theory is required to relate 
$\Delta \mu$ of such theories with
$\Delta \mu$ of the reservoirs, by which $G$ is defined.

An advantage of the present method is that 
we do not need to find the relation between 
$\Delta \phi_{\rm ext}$ and $\Delta \mu$ because
$\langle I \rangle$ is directly calculated 
as a function of $\Delta \mu$.
Another advantage is that 
it is applicable to NEN and nonlinear responses
because nonequilibrium steady state is directly obtained.

\subsection{Conductance of the 1d Fermi liquid}
\label{sec_FL}

It is generally believed that a
1d interacting electron system is not the Fermi liquid (FL) 
\cite{nozieres}, 
but the Tomonaga-Luttinger liquid (TLL)
\cite{tomonaga,luttinger,haldane,KY}.
For this reason, many papers on 1d systems 
\cite{AR,KF,FN,OF,maslov,pono,safi,oreg96}
use the word FL to indicate {\it non-interacting} electrons, i.e., 
a Fermi {\it gas}.
However, we do not use such a misleading terminology;
by a FL we mean {\it interacting} quasi-particles.
Since the backflow is induced by the interaction \cite{nozieres}, 
the Landauer's argument of non-interacting particles \cite{LB}
cannot be applied to a FL.
On the other hand, 
real systems have finite length and finite 
intersubband energies, 
in contradiction to the assumptions of the TLL.
Hence, some real systems
might be well described as a FL.
Therefore, $G$ of a FL is 
non-trivial and interesting \cite{nontrivial}.
Furthermore, we will show in section \ref{sec_nmf} that 
the results for the FL suggest very important 
phenomena that is characteristic to nonequilibrium states
of inhomogeneous systems.
Note also that the following calculations look similar to 
the derivation of fundamental relations in the theory of
the FL \cite{nozieres}.
However, $G$ of mesoscopic conductors 
was not calculated in such calculations.
The most important point to evaluate $G$ is 
to find the nonequilibrium steady state.

We find the nonequilibrium steady state
using a thermodynamical argument as follows:
In the reservoirs, electrons behave as a 2d or 3d (depending 
on the thickness of the reservoir regions) FL.
Since we have assumed that $u^{\rm c}$ is smooth and slowly-varying, 
a 2d or 3d quasi-particle in a reservoir, 
{\it together with its backflow}, 
can evolve adiabatically into a 1d quasi-particle 
and its backflow in the QWR, without reflection.
In this adiabatic evolution, 
the quasi-particle mass $m^*$ and the Landau parameters $f$
also evolve adiabatically, and the energy is conserved.
Therefore, 
quasi-particles with $\varepsilon (k >0) \leq \mu_{\rm L}$ 
are injected from the left reservoir.
Here, $\varepsilon$ is the quasi-particle energy;
\begin{equation}
\varepsilon(k)
=
\frac{\hbar^2 k^2}{2 m^*} 
+
\frac{\hbar}{\cal L}
\sum_{k'} f(k, k') \delta n(k'),
\label{qpe}\end{equation}
where 
$
\delta n(k)
\equiv
n(k) - \Theta(|k| \leq k_{\rm F})
$, with $n(k)$ being the quasi-particle distribution.
The last term of this expression represents energy correction 
by interactions among quasi-particles \cite{nozieres}.
On the other hand, a quasi-hole below $\mu_{\rm L}$ 
should not be injected because otherwise the recombination 
of a quasi-particle with the quasi-hole would produce excess entropy, 
in contradiction with the principle of minimum entropy 
production. 
Similarly, 
quasi-particles with $\varepsilon (k <0) \leq \mu_{\rm R}$ 
are injected from the right reservoir, 
with no quasi-holes are injected below $\mu_{\rm R}$.
Therefore, 
the nonequilibrium steady state under a finite 
$\Delta \mu = \mu_{\rm L} - \mu_{\rm R}$ should be the 
``shifted Fermi state", 
in which quasi-particle states with 
$\varepsilon(k \geq 0) \leq \mu_{\rm L}$ 
and
$\varepsilon(k<0) \leq \mu_{\rm R}$ 
are all occupied.
Hence, the right- (left-) going quasi-particles 
have the chemical potential $\mu_+ = \mu_{\rm L}$
($\mu_+ = \mu_{\rm R}$).
Considering also the charge neutrality, we can write 
the distribution function as
\begin{equation}
n(k) = \Theta(|k-q| \leq k_{\rm F}).
\label{sfd}
\end{equation}
Then, Eq.\ (\ref{qpe}) yields
\begin{equation}
\mu_{\pm}
=
\frac{\hbar^2 k_{\rm F}^2}{2 m^*} 
\pm 
\hbar q \left[
\frac{\hbar k_{\rm F}}{m^*}
+ 
\frac{f_{++} - f_{+-}}{2 \pi}
\right],
\label{muFL}\end{equation}
where 
$f_{++} \equiv f(k_{\rm F},k_{\rm F})$ and 
$f_{+-} \equiv f(k_{\rm F},-k_{\rm F}) = f(-k_{\rm F},k_{\rm F})$.
Hence, 
\begin{equation}
\Delta \mu
=
2 \hbar q \left[
\frac{\hbar k_{\rm F}}{m^*}
+ 
\frac{f_{++} - f_{+-}}{2 \pi}
\right].
\label{DmuFL}
\end{equation}
On the other hand, considering the spin degeneracy, 
$\langle I \rangle$ 
is calculated as
\begin{equation}
\langle I \rangle 
= 2 e {q \over \pi}
\left[
\frac{\hbar k_{\rm F}}{m^*}
 +
\frac{f_{++} - f_{+-}}{2 \pi}
\right].
\label{IFL}\end{equation}
Here, the $f_{+ \pm}$-dependent terms represent the backflow.
Since the same factor appears in Eq.\ (\ref{DmuFL}), 
we find that the conductance is independent of 
$m^*$ and $f_{+ \pm}$;
\begin{equation}
G \equiv 
\frac{\langle I \rangle}{\Delta \mu/e}
=
\frac{e^2}{\pi \hbar}.
\label{GFL}\end{equation}
Since we have identified the nonequilibrium steady state, 
we can calculate not only $G$ but also 
other nonequilibrium properties such as the NEN 
\cite{s96}.

It is instructive to represent Eqs.\ (\ref{muFL})-(\ref{IFL}) 
in terms of the bare parameters.
As in the case of 3d Fermi liquid \cite{nozieres},
we can show that \cite{note_mass}
\begin{equation}
\frac{\hbar k_{\rm F}}{m}
=
\frac{\hbar k_{\rm F}}{m^*}
+ 
\frac{f_{++} - f_{+-}}{2 \pi}.
\label{mass}\end{equation}
Hence, we can rewrite Eq.\ (\ref{IFL}) as
$
\langle I \rangle 
= 2 e (q / \pi)
(\hbar k_{\rm F} /m)
$.
Therefore, 
quasi particles (whose group velocity is $\hbar k_{\rm F} /m^*$)
plus their backflows carry exactly the same current as
the bare particles, for the same $q$, i.e., 
for the same shifted Fermi distribution.
On the other hand,
Eq.\ (\ref{muFL}) is rewritten as
$
\mu_{\pm}
=
\hbar^2 k_{\rm F}^2/2 m^* 
\pm 
\hbar^2 q k_{\rm F}/m
$.
Although $\mu_{\pm}$ $\neq$ [$\mu_{\pm}$ of bare particles],
$\Delta \mu$ $=$ [$\Delta \mu$ of bare particles]
for the same $q$.
These facts result in the independence of $G$ on the Landau 
parameters.

\subsection{Conductance of the Tomonaga-Luttinger liquid}
\label{sec_TLL}

We now consider a clean TLL \cite{s96}.
The low-energy dynamics of a TLL
is described by the charge ($\rho$) 
and spin ($\sigma$) excitations
(whose quantum numbers are $N_q^\rho$ and $N_q^\sigma$, 
respectively, where $q \neq 0$ denotes the wavenumber), 
and the zero modes 
(quantum numbers $N_\pm^\rho$, $N_\pm^\sigma$)
\cite{tomonaga,luttinger,haldane,KY}.
The eigenenergy is given by
\begin{equation}
E 
= 
\sum_{\nu = \rho, \sigma}
 v^\nu
\sum_q \hbar |q| N_q^\nu
+ \frac{\pi \hbar}{2 {\cal L}}
\sum_{\nu = \rho, \sigma}
[v_N^\nu(N_+^\nu + N_-^\nu)^2 + v_J^\nu (N_+^\nu - N_-^\nu)^2],
\label{ETLL}
\end{equation}
where
$v_N^\nu = v^\nu/K_\nu$ and 
$v_J^\nu = K_\nu v^\nu$ ($\nu = \rho, \sigma$).
Here, the parameters $v^\nu$ and $K_\nu$ 
are renormalized by the $e$-$e$ interactions
(except that $K_\sigma=1$ by the SU(2) symmetry).
The DC current is given by 
\begin{equation}
\langle I \rangle
=
2 e 
 v_J^\rho (N_+^\rho - N_-^\rho)/{\cal L}.
\label{ITLL}
\end{equation}
We apply a thermodynamical argument
to find the nonequilibrium steady state.
Unlike the FL case, there is no adiabatic continuity between the 
TLL in the QWR and the FL in the reservoirs.
We therefore argue differently:
In the linear response regime
the steady state must be the state with the minimum energy
among states which satisfy given external conditions.
Otherwise, the system would be unstable and
would evolve into a state with lower energy.
For our purpose, it is convenient to
take the value of $\langle I \rangle$ as the given external condition.
Then, from Eqs.\ (\ref{ETLL}) and (\ref{ITLL}), 
we find that
the steady state should be the state with 
$N_q^\rho = N_q^\sigma = 0$ (for all $q$), 
$N_+^\rho + N_-^\rho = 0$, 
$N_+^\sigma + N_-^\sigma 
= N_+^\sigma - N_-^\sigma =0$, and
$N_+^\rho - N_-^\rho > 0$.
This state may be called the ``shifted Fermi state" of the TLL.
Furthermore,
in the steady state, 
electrons in the left reservoir
and right-going electrons in the TLL
should be in the ``chemical equilibrium", 
in which electrons in the FL phase are 
transformed into right-going electrons in the TLL phase
at a constant rate.
Therefore, their chemical potentials should be equal \cite{note_mu};
\begin{equation}
\mu_{\rm L,R} = \mu^\rho_{+,-} 
\equiv \frac{\partial E}{\partial N_{+,-}^\rho}
= {\pi \hbar \over {\cal L}}[v_N^\rho(N_+^\rho + N_-^\rho) 
\pm v_J^\rho(N_+^\rho - N_-^\rho)],
\label{muisequalL}\end{equation}
where we have used Eq.\ (\ref{ETLL}).
Hence, 
\begin{equation}
\Delta \mu = \mu_{\rm L} - \mu_{\rm R}
=
{2 \pi \hbar \over {\cal L}}
v_J^\rho(N_+^\rho - N_-^\rho).
\label{DmuTLL}\end{equation}
By dividing Eq.\ (\ref{ITLL}) by this expression, 
we obtain the same result for $G$ as Eq.\ (\ref{GFL}), 
in agreement with experiment \cite{tarucha}.

Since we have identified the nonequilibrium steady state, 
we can calculate not only $G$ but also 
other nonequilibrium properties such as the NEN 
\cite{s96}.

\section{Projection theory}
\label{sec_proj}

Although we have successfully found the nonequilibrium steady state
of interacting electrons in section \ref{sec_combined},
a possible objection against the formulation may be 
that the theory is rather intuitive.
In this section, 
we present a full statistical-mechanical theory \cite{sm98,ks}, 
which is free from such an objection.
In this theory, 
we start from the Hamiltonian 
of 3d interacting electrons confined in the composite system of
the QWR and reservoirs, Fig.\ \ref{fig1}. 
This original system 
is projected onto an effective 1d system, 
and the equation of motion for the reduced density operator 
of the 1d system is derived.
From this equation, we can find 
the nonequilibrium steady state as a function of $\Delta \mu$ 
between the reservoirs.
This allows us to evaluate various nonequilibrium properties.

\subsection{Decomposition of the 3d electron field}
\label{sec_decompose}

We start from the 3d electron field $\efn$
subject to a confining potential $u^{\rm c}(\r)$
(which defines the QWR and two reservoirs connected to it), 
impurity potential $u^{\rm i}(\r)$
(whose average $\overline{u^{\rm i}}$ is absorbed 
in $u^{\rm c}(\r)$, hence $\overline{u^{\rm i}}=0$), 
external electrostatic potential $\phi_{\rm ext}(\r)$,
and the 
e-e interaction of equal strength $v(\r - \r')$ in all regions \cite{sm98}:
\begin{eqnarray}
\hat H 
&=&
\int d^3r \
\efd \left[
- \frac{\hbar^2}{2 m} \nabla^2 
+ u^{\rm c}(\r)+u^{\rm i}(\r) + e \phi_{\rm ext}(\r)
\right] \efn
\nonumber\\
& & +
\frac{1}{2}
\int d^3r \int d^3r' \ 
\hat \rho(\r)v(\r - \r') \hat \rho(\r'),
\label{H3d}\end{eqnarray}
where $\hat \rho(\r)$ is the charge density.
We will find the nonequilibrium steady state for $\Delta \mu>0$.
For this state, $\langle \hat \rho(\r) \rangle \neq 0$, which 
gives rise to a long-range force. 
We extract it as 
the renormalization of the electrostatic potential 
\begin{equation}
e \phi(\r)
=
e \phi_{\rm ext}(\r) 
+
\int d^3r' \ 
v(\r - \r') \langle \hat \rho(\r') \rangle,
\label{ren_phi}\end{equation}
and a c-number $V_{\rm av}$.
Namely, $\hat H$ is recast in terms of 
$
\delta \hat \rho(\r)
\equiv
\hat \rho(\r) - \langle \hat \rho(\r) \rangle
$
as
\begin{eqnarray}
\hat H 
&=&
\int d^3r \
\efd 
\left[
- \frac{\hbar^2}{2 m} \nabla^2 
+ u^{\rm c}(\r)+u^{\rm i}(\r) + e \phi(\r)
\right] 
\efn
\nonumber\\
& & +
\frac{1}{2}
\int d^3r \int d^3r' \ 
\delta \hat \rho(\r)v(\r - \r') \delta \hat \rho(\r')
+ V_{\rm av}.
\label{H3d_recast}\end{eqnarray}
%

To decompose $\efn$, we consider the single-body part of $\hat H$.
Recall that $u^{\rm c}(\r)$ is assumed to be smooth and slowly-varying,
to avoid undesirable reflections at the QWR-reservoir boundaries.
In this case, the single-body Schr\"odinger equation
\begin{equation}
\left[
- \frac{\hbar^2}{2 m} \nabla^2 
+ u^{\rm c}(\r)+u^{\rm i}(\r) + e \phi(\r)
\right] 
\varphi(\r)
=
\varepsilon 
\varphi(\r)
\end{equation}
has solutions 
 that propagate through the QWR having the energy  
$\varepsilon \simeq \varepsilon_{\rm F}$ \cite{adiabatic};
\begin{equation}
\varphi_{k}(\r) \simeq 
{1 \over \sqrt{\cal L}}
\exp \left[ i \int_0^x K_{k}(x) dx \right]
\varphi^\bot(y,z;x).
\end{equation}
Here, $\varphi^\bot(y,z;x)$ 
is the wavefunction of the lowest subband at $x$, 
representing the confinement in the lateral ($yz$)
directions, 
and ${\cal L}$ is the normalization length in the $x$ direction.
All the other modes are denoted by $\varphi_\nu(\r)$, which 
includes solutions that are localized in either reservoir, 
and extended solutions whose $\varepsilon$ are not close to 
$\varepsilon_{\rm F}$.
Since any function of $\r$ 
can be expanded in terms of $\varphi_k(\r)$'s and 
$\varphi_\nu(\r)$'s, so is the $\r$ dependence of the 
electron field operator;
\begin{equation}
\hat \psi(\r)
=
\sum_k \hat c_k \varphi_k(\r)
+
\sum_\nu \hat d_\nu \varphi_\nu(\r)
\equiv
\varphi^\bot(y,z;x) \ofn
+
\hat \psi_{\rm R}(\r).
\end{equation}
The 3d electron field has thus been decomposed into
the 1d field and
the 3d field $\hat \psi_{\rm R}(\r)$, which we call the 
``reservoir field.'' 
It can be decomposed into
the low-energy components $\hat \psi_{\rm R_{\rm L}}$ and 
$\hat \psi_{\rm R_{\rm R}}$
(which are localized in the left and right reservoirs, 
R$_{\rm L}$ and R$_{\rm R}$, respectively)
and 
the high-energy component $\hat \psi_{\rm R_H}$ as
\begin{equation}
\hat \psi_{\rm R} = 
\hat \psi_{\rm R_{\rm L}} + \hat \psi_{\rm R_{\rm R}} + \hat \psi_{\rm R_H}.
\end{equation}
For low-energy phenomena, 
we can take
$\hat \psi_{\rm R} = \hat \psi_{\rm R_{\rm L}} + \hat \psi_{\rm R_{\rm R}}$.



\subsection{Hamiltonian for the 1d and the reservoir fields}
\label{sec_H}

By expressing $\hat H$ 
in terms of $\hpn_1$ and $\hpn_{\rm R}$, 
we obtain 
the single-body part of $\hat \psi_1$ (denoted by $\hat H_{1}^0$), 
the $\hat \psi_1$-$\hat \psi_1$ interaction ($\hat V^0_{11}$),
the $\hat \psi_1$-$\hat \psi_{\rm R}$ interaction ($\hat V_{\rm 1R}^0$),
the $\hat \psi_{\rm R}$-$\hat \psi_{\rm R}$ interaction 
($\hat V_{\rm RR}^0$), 
and the single-body part of $\hat \psi_{\rm R}$ ($\hat H_{\rm R}^0$)
 \cite{sm98}.
By the screening effect of $\hat V_{\rm RR}^0$, $\hat V_{\rm 1R}^0$ is 
renormalized as the screened interaction $\hat V_{\rm 1R}$.
Similarly, by the screening effect of $\hat V_{\rm 1R}$, $\hat V_{11}^0$ is 
renormalized as the screened interaction $\hat V_{11}$.
We therefore recast 
$\hat V_{11}^0 + \hat V_{\rm 1R}^0 + \hat V_{\rm RR}^0$
as 
$\hat V_{11} + \hat V_{\rm 1R} + \hat V_{\rm RR}$, 
where 
$\hat V_{\rm 1R}$ and $\hat V_{\rm RR}$ have no screening effects on 
$\hat V_{11}$ and $\hat V_{\rm 1R}$.
In this way, $\hat H$ is decomposed as
\begin{equation}
\hat H = \hat H_1 + \hat V_{\rm 1R} + \hat H_{\rm R},
\label{H}\end{equation}
where 
$\hat H_1 \equiv \hat H_{1}^0 + \hat V_{11}$
is the Hamiltonian for $\ofn$, 
$\hat H_{\rm R} \equiv \hat H_{\rm R}^0 + \hat V_{\rm RR}$
is the one for $\rfn$,
and 
$\hat V_{\rm 1R}$ is the interaction between 
$\ofn$ and $\rfn$.
In particular, 
$\hat H_1$ is evaluated as
\begin{eqnarray}
\hat H_1
&=&
\int dx \
\ofd \left[ 
- {\hbar^2 \over 2m} {\partial^2 \over \partial x^2} 
+ u^\bot(x) +u^{\rm i}_1(x) 
\right] \ofn
\nonumber\\
& & +
\frac{1}{2} \int dx \int dx' \
\delta \hat \rho_1(x,t)v_{11}(x,x') \delta \hat \rho_1(x',t).
\end{eqnarray}
Here,
$u^\bot$ is the subband energy \cite{sm98}, 
$\delta \hat \rho_1(x,t) \equiv \ofd \ofn - \bra \ofd \ofn \ket$
is the density fluctuation of the 1d field, 
and 
\begin{eqnarray}
u^{\rm i}_1(x)
&\equiv&
\int \! \int dy dz \ |\varphi^\bot(y,z;x)|^2 \ u^{\rm i}(\r),
\label{ui}\\
v_{11}(x,x')
&\equiv&
\int \! \int \! \int \! \int \! dy dz dy' dz' \ 
|\varphi^\bot(y,z;x)|^2 
v^{\rm sc}(\r, \r') \ 
|\varphi^\bot(y',z';x')|^2
\end{eqnarray}
are the impurity and two-body potentials for the 1d field,
where $v^{\rm sc}$ is the screened two-body potential for $\hpn$.

\begin{figure}[b]
\begin{center}
\includegraphics[width=.9\textwidth]{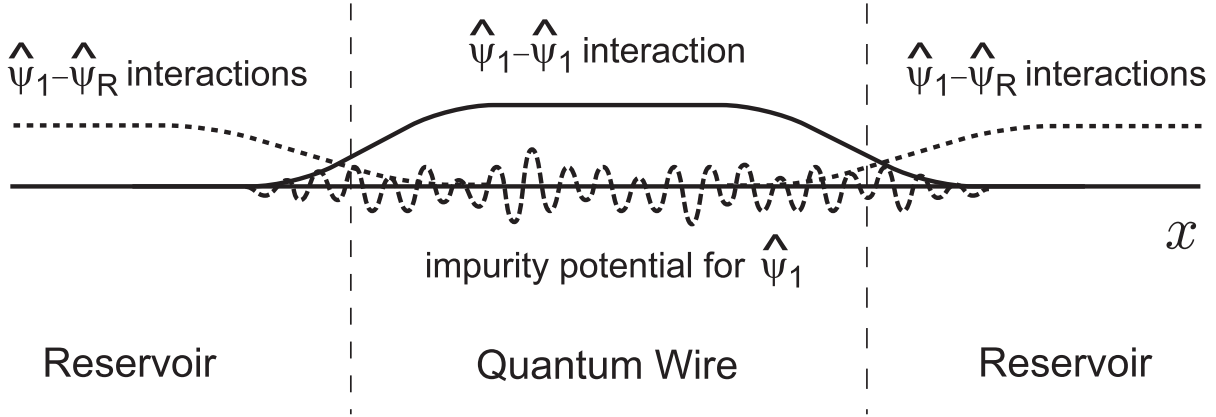}
\end{center}
\caption[]{
Schematic diagram of the strengths of scatterings 
of the 1d field. 
}
\label{effective_int}
\end{figure}

It is seen that 
$u^{\rm i}_1(x)$ is 
the average of the random potential
$u^{\rm i}_1(\r)$ over the lateral wavefunction $\varphi^\bot$, 
which, as a function of $y$, 
is localized in a region of width $\sim W(x)$ for each $x$.
Here, 
$W(x)$ denotes the width of the region in which electrons 
are confined  (Fig.\ \ref{fig1}).
From these observations, we can show that
\begin{eqnarray}
u^{\rm i}_1(x) &\sim& u^{\rm i}(\r) 
\quad \mbox{for $x \in$ QWR}, 
\\
|u^{\rm i}_1(x)| &\propto& 1/\sqrt{W(x)}
\quad \mbox{for $x \in$ a reservoir}.
\label{ui_2}\end{eqnarray}
In a similar manner, 
the two-body potential for the 1d field behaves as
\begin{eqnarray}
v_{11}(x,x') 
&\lesssim&
v^{\rm sc}([|x-x'|^2+W(0)^2]^{1/2})
\quad \mbox{for $x$ or $x' \in$ QWR}, 
\\
v_{11}(x,x') 
&\sim&
(r^{\rm sc}/ W) \overline{v^{\rm sc}},
\quad \mbox{for $x \sim x'$ $\in$ a reservoir}.
\label{v11_2}\end{eqnarray}
where $r^{\rm sc}$ denotes the range of $v^{\rm sc}$, 
and $\overline{v^{\rm sc}}$ the average of $v^{\rm sc}$
in the region $|\r-\r'| \lesssim r^{\rm sc}$.
%
We now assume that the width of the reservoirs is very large;
\begin{equation}
W(x) \to \infty \mbox{ as } x \to \pm \infty.
\end{equation}
Then, Eqs.\ (\ref{ui_2})-(\ref{v11_2}) yield
$u^{\rm i}_1(x)\to 0$ as $x \to \pm \infty$, 
and 
$v_{11}(x,x') \to 0$ as $x$ or $x'$ $\to \pm \infty$.
Namely, 
$\hat H_1$ represents interacting 
$\ofn$ field that gets free as $x \to \pm \infty$.
On the other hand, the interaction 
$\hat V_{\rm 1R}$ between $\ofn$ and $\rfn$
becomes stronger in the reservoir regions, 
whereas it is negligible in the QWR
because at low energies $\rfn$ does not penetrate into the QWR.
Therefore, the 1d field $\ofn$ is subject to 
different scatterings 
in different regions of $x$:
In the QWR, $\hpn_1$ is scattered by 
the $\hpn_1$-$\hpn_1$ interaction and the impurity potentials, 
whereas $\hpn_1$ is excited and attenuated by 
the reservoir field in the reservoir regions
through the $\hpn_1$-$\hpn_{\rm R}$ interactions (Fig.\ \ref{effective_int}).

\subsection{Equation of motion for the reduced density operator 
}
\label{sec_eqm}

We have successfully rewritten the Hamiltonian $\hat H$ 
in terms of $\hpn_1$ and $\hpn_{\rm R}$.
We must go one step further because 
$\hat H = \hat H_1 + \hat V_{\rm 1R} + \hat H_{\rm R}$ describes
very complicated dynamics, and thus
the von Neumann equation for the density operator $\hat \zeta$,
\begin{equation}
i \hbar \frac{\partial}{\partial t}
\hat \zeta (t) 
=
\left[
\hat H_1 + \hat V_{\rm 1R} + \hat H_{\rm R}, 
\hat \zeta (t) 
\right]
\end{equation}
is impossible to solve.
{\em This unsolvability guarantees the thermodynamical 
stability (mixing property) of the total system} \cite{ott,nakano,dset}.
We turn this fact to our own advantage, and  
reduce the theory to a tractable one \cite{sm98,ks}.
The basic idea is as follows:
$\hat H_1$ describes the 1d correlated electrons, and is most
important. Hence, it should be given a full quantum-mechanical 
treatment.
Concerning $\hat V_{\rm 1R}$, on the other hand, 
multiple interactions by $\hat V_{\rm 1R}$ seem unimportant.
Hence, we may treat it by a second-order perturbation theory \cite{2nd}.
For $\hat H_{\rm R}$, it describes the 2d or 3d interacting electrons, 
for which many properties are well known, 
and we can utilize the established results.
Moreover, since the reservoirs are large, 
we can assume the {\em local equilibrium}:
both reservoirs are in their equilibrium states with 
the chemical potentials $\mu_{\rm L}$ and $\mu_{\rm R}$, respectively.
We denote the reduced density operator of the reservoir field
for this local equilibrium state by $\hat \zeta_{\rm R}$. 

From these observations, we may 
project out the reservoir field $\hpn_{\rm R}$ as follows.
Consider the reduced density operator for the 1d field;
$ 
\hat \zeta_1(t) \equiv
{\rm Tr}_{\rm R} [
\hat \zeta(t)
].
$ 
Up to the second order in $\hat V_{\rm 1R}$ \cite{2nd}, 
the equation of motion of $\hat \zeta_1$ 
in the interaction picture of $\hat H_1 + \hat H_{\rm R}$,
is evaluated as
\begin{equation}
\frac{\partial}{\partial t} 
\hat \zeta_1(t)
=
\frac{-1}{\hbar^2}
\int_{-\infty}^{t} dt'
{\rm Tr}_{\rm R} \left(
\left[
\hat V_{\rm 1R}(t), 
\left[
\hat V_{\rm 1R}(t'), \hat \zeta_{\rm R} \hat \zeta_1(t)
\right]
\right]
\right),
\label{master}\end{equation}
where 
we have used the fact that 
$\hat \zeta_1(t)$ in the interaction 
picture varies only slowly, so that 
$\hat \zeta_1(t') \simeq \hat \zeta_1(t)$ in the 
correlation time of  
$\langle \hat V_{\rm 1R}(t) \hat V_{\rm 1R}(t') \rangle$ \cite{precise}.
This equation represents that 
$\hat \zeta_1$ is driven by two reservoirs,
which have different chemical potentials $\mu_{\rm L}$ 
and $\mu_{\rm R}$, through the $\hpn_1$-$\hpn_{\rm R}$ interaction.
Since the trace is taken over the reservoir field, 
Eq.\ (\ref{master}) is a closed equation for $\hpn_1$ and
$\hat \zeta_1$.
Its steady solution 
represents the 
nonequilibrium steady state of the 1d field driven by 
the reservoirs.

\subsection{Current of the 1d field}
\label{sec_1dcurrent}

\begin{figure}[t]
\begin{center}
\includegraphics[width=.9\textwidth]{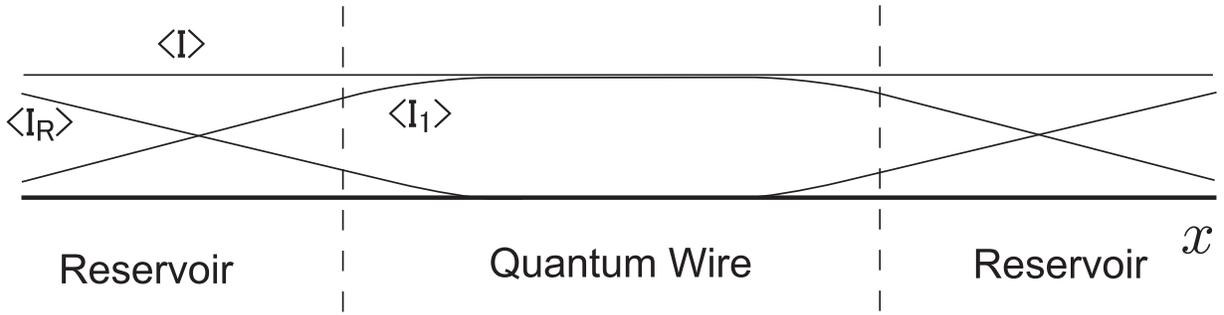}
\end{center}
\caption[]{
Schematic plots of the expectation values 
$\langle I \rangle$, 
$\langle I_1 \rangle$,
and 
$\langle I_{\rm R} \rangle$,
of the currents carried by $\hpn$, 
$\hpn_1$, and $\hpn_{\rm R}$, 
respectively.
}
\label{fig_current}
\end{figure}

We now turn to observables.
We are most interested in 
the {\it total} current $\hat I$ which is given by 
$ 
\hat I(x,t) \equiv \int \int dy dz 
\hat J_x (\r,t),
$ 
where $\hat J_x$ denotes the $x$ component of the current density.
Note that $\hat I$ is different from 
the current of the 1d field defined by \cite{sm98}
\begin{equation}
\hat I_1(x,t) 
\equiv
{e \over 2 m} \left[
\hpd_1(x,t)
\left\{ {\hbar \over i}{\partial \over \partial x}
\hpn_1(x,t)
\right\} + {\rm h.c.}
\right].
\label{I1}\end{equation}
The expectation values of $\hat I$, 
$\hat I_1$, and 
$\hat I_{\rm R}$ (the current carried by the reservoir field)
are schematically plotted in Fig.\ \ref{fig_current}, 
which shows that 
$\langle I \rangle$ is mainly carried by 
$\langle I_1 \rangle$ in the QWR and by 
$\langle I_{\rm R} \rangle$ in the reservoirs, respectively.
The transformation between $\hat I_1$ and  
$\hat I_{\rm R}$ is caused by $\hat V_{\rm 1R}$, 
and thus 
$\hat I_1$ is {\it not conserved}:
$
\px \hat I_1 + \pt \hat \rho_1 \neq 0
$.
At first sight, these facts might seem to cause difficulties in 
calculating $\langle I \rangle$ and $\langle \delta I^2 \rangle$ 
from $\hat \zeta_1$.
Fortunately, however, we can show that  
for any (nonequilibrium) steady state
$\langle I \rangle$ and
$\langle \delta I^2 \rangle^{\omega \simeq 0}$
are independent of $x$, and 
that 
\begin{eqnarray}
\langle I \rangle 
=
\langle I_1 \rangle
&&
\quad \mbox{at $x \simeq 0$}
\\
\langle \delta I^2 \rangle^{\omega \simeq 0} 
=
\langle \delta I^2_1 \rangle^{\omega \simeq 0} 
&&
\quad \mbox{at $x \simeq 0$}
\label{IisI1}\end{eqnarray}
where $x=0$ corresponds to the center of the QWR.
Therefore, 
to calculate $\langle I \rangle$ and $\langle \delta I^2 \rangle$,
it is sufficient to calculate 
$\langle I_1 \rangle$ and
$\langle \delta I^2_1 \rangle^{\omega \simeq 0}$
at $x \simeq 0$, 
which can be calculated from $\hat \zeta_1$.
Therefore, 
we have successfully reduced
the 3d problem, Eq.\ (\ref{H3d}), into 
the effective 1d problem,
Eqs.\ (\ref{H}), (\ref{master}) and (\ref{I1}).

Actual calculations can be conveniently performed as follows.
Although $\langle I \rangle$ and $\langle \delta I^2 \rangle$
have both low- and high-frequency components,
we are only interested in the low-frequency components, 
which are denoted by $\bar{I}(t)$ and $\delta \bar{I}^2(t)$, 
respectively.
They are given by
\begin{equation}
\bar{I}(t)
=
\int_{-\infty}^{\infty} dt'
f(t'-t) {\rm Tr} [ \hat I(t') \hat \zeta(t') ],
\label{bI}\end{equation}
and similarly for $\delta \bar{I}^2(t)$.
Here, $f(t'-t)$ is a filter function that 
is finite only in the region $t-\tau/2 \lesssim t' \lesssim t + \tau/2$,
where $1/\tau \simeq$ the highest frequency of interest.
From Eqs.\ (\ref{H}), (\ref{master})-(\ref{bI}), 
we can construct
the equations for $\bar{I}(t)$ and $\delta \bar{I}^2(t)$.
They can be solved more easily than the equation for $\hat \zeta_1$, 
and the solutions fully describe the low-frequency behaviors
of $\langle I \rangle$ and $\langle \delta I^2 \rangle$.

In the following, we present the results for $\bar{I}(t)$ for 
the case of impurity scatterings and for the the case of 
$e$-$e$ interactions.

\subsection{Application of the projection theory to 
the case where impurity scatterings are present in all regions}
\label{sec_proj_impurity}

When electrons are scattered by impurities (one-body potentials)
in all regions including reservoirs, 
whereas many-body scatterings are negligible, 
the $\hpn_1$-$\hpn_{\rm R}$ interaction is given by
\begin{equation}
\hat V_{\rm 1R}
=
\int d^3 r
\hat \psi_1^\dagger (x)
\varphi^{\bot *}(y,z;x)
u^{\rm i}(\r) \rfn
+{\rm h.c.}
\end{equation}
If we put 
$
\hat Y_\alpha(\r)
\equiv
\varphi^{\bot *}(y,z;x)
u^{\rm i}(\r) \hat \psi_{\rm R_\alpha}(\r)
$ ($\alpha=$ L, R),
then Eq.\ (\ref{master}) becomes
\begin{eqnarray}
\frac{\partial}{\partial t} \hat \zeta_1 (t)
&=&
\frac{-1}{\hbar^2}
\sum_\alpha
\int_{-\infty}^t \!\!\!\!\!\! dt' \!
\int \!\! d^3 r \! 
\int \!\! d^3r' 
\biggr\{
\left\langle \hat Y_\alpha^\dagger (\r,t) 
\hat Y_\alpha(\r',t') \right\rangle_\alpha
\left[
\hat \psi_1(x,t),
\hat \psi_1^\dagger(x',t')
\hat \zeta_1(t)
\right]
\nonumber\\
&+&
\left\langle \hat Y_\alpha(\r,t) 
\hat Y_\alpha^\dagger (\r',t') \right\rangle_\alpha
\left[
\hat \psi_1^\dagger(x,t),
\hat \psi_1(x',t')
\hat \zeta_1(t)
\right]
\biggr\}
+{\rm h.c.}
\label{master2}\end{eqnarray}
where
$\langle \cdots \rangle_\alpha$ denotes the expectation value 
for the equilibrium state of reservoir R$_\alpha$, 
which has the chemical potential $\mu_\alpha$.
After careful calculations using Eqs.\ (\ref{I1})-(\ref{bI}), 
and considering the spin degeneracy,
we find \cite{ks}
\begin{equation}
\frac{d}{dt} 
\bar I(t)
=
- \gamma 
\left[
\bar I(t)-
\bar I_{\rm steady}
\label{EQMofI}\right],
\end{equation}
where
$
\bar I_{\rm steady}
\equiv
(e/\pi \hbar) T \Delta \mu
$, and 
$ 
\gamma 
\simeq
(2 \pi / \hbar) n_{\rm imp} |u^{\rm i}|^2 {\cal D}_{\rm F} 
$. 
Here, $n_{\rm imp}$ is the impurity density, 
$u^{\rm i}$ denotes the potential of an impurity 
($u^{\rm i}(\r) = \sum_{\ell} u^{\rm i} \delta (\r-\r_{\ell})$),
and ${\cal D}_{\rm F}$ is the density of states per unit volume, 
${\cal D}(\mu_{\rm L}) \simeq {\cal D}(\mu_{\rm R}) \equiv {\cal D}_{\rm F}$.
It is seen that $\bar I$ approaches 
$
\bar I_{\rm steady}
$
as $t \to \infty$.
Therefore, the DC conductance is given by
\begin{equation}
G = T \frac{e^2}{\pi \hbar},
\end{equation}
in agreement with
the Landauer-B\"uttiker formula \cite{LB}. 
Moreover, we find that 
{\it the steady state is stable}:
For any (small) deviation from the steady state, 
$\bar I$
relaxes to the value $\bar I_{\rm steady}$, 
with the relaxation constant $\gamma$.

\subsection{Application of the projection theory to 
the case where $e$-$e$ scatterings are present in all regions}
\label{sec_proj_ee}

When the $e$-$e$ interaction is important
in all regions including reservoirs, 
whereas impurity scatterings are negligible, 
we find that the most relevant term of the $\hpn_1$-$\hpn_{\rm R}$ 
interaction is given by \cite{ks}
\begin{equation}
\hat V_{\rm 1R}
=
\int \!\! d^3 r \int \!\! d^3r' 
\hat \psi_1^\dagger (x) \varphi^{\perp *}(y,z;x)
\hat \psi_{\rm R}(\r) v^{\rm sc}(\r,\r')
 \hat \psi_{\rm R}^\dagger(\r')
 \hat \psi_{\rm R}(\r')
+ {\rm h.c.} 
\end{equation}
By this interaction, 
an electron is scattered into the QWR through 
the collision of two electrons in a reservoir, or, 
an electron in the QWR is absorbed in a reservoir.
By putting
$
\hat Y_\alpha(\r)
\equiv
\int \!\! d^3r' 
\varphi^{\perp *}(y,z;x)
\hat \psi_{\rm R_\alpha}(\r) v^{\rm sc}(\r,\r')
 \hat \psi_{\rm R_\alpha}^\dagger(\r')
 \hat \psi_{\rm R_\alpha}(\r')
$
(for $\alpha =$ L, R), 
which differs from $\hat Y_\alpha$ of the impurity-scattering case, 
we obtain the equation of motion for $\hat \zeta_1$
in the same form as Eq.\ (\ref{master2}).
To derive the equation of motion for 
$
\bar I
$
from that equation, 
we need to calculate the correlation functions
in the reservoirs, 
$
\langle \hat Y_\alpha(t) 
\hat Y_\alpha^\dagger(t') \rangle_{\alpha}
$
and
$
\langle \hat Y_\alpha^\dagger(t) 
\hat Y_\alpha(t') \rangle_{\alpha}
$.
We can easily calculate them using well-known results for 
the 2D or 3D FL because the reservoir electrons are believed to be 
the 2D or 3D FL.
We also need correlation functions of the 1d field. 
They are quite different depending 
on the nature (FL or TLL) of the 
electrons in the QWR.

In the case where $\hpn_1$ behaves as a 1d FL, 
we obtain the equation for 
$\bar{I}(t)$
in the same form as Eq.\ (\ref{EQMofI}),
but now $\gamma$ is a function of the $e$-$e$ interaction 
parameters, and
\begin{equation}
\bar{I}_{\rm steady}
=
2 \frac{e}{\cal L}
\sum_{k>0}
\frac{\hbar k}{m}
[
\Theta (\mu_{\rm L} - \varepsilon(k))
-
\Theta (\mu_{\rm R} - \varepsilon(-k))
].
\end{equation}
Here, 
$m$ is the {\it bare} mass, 
$\Theta$ is the step function,
and 
$
\varepsilon(\pm k)
$
denotes the 1d quasi-particle energy, Eq.\ (\ref{qpe}),  
{\it in the shifted Fermi state}.
Note that if we simply took 
$
\varepsilon(\pm k) = \hbar^2 k^2 / 2 m^*
$,
then 
$
\bar I_{\rm steady}
=
(m^*/m)(e/\pi \hbar) \Delta \mu
$,
hence the conductance would be renormalized 
by the factor $m^*/m$.
However, 
the correct expression (\ref{qpe})
shows that 
$
\varepsilon(\pm k)
$
{\it are modified in the presence of a finite current},  
and the correction terms are proportional 
to $q \propto \bar I$. 
As a result, the injection of an electron becomes easier or harder 
as compared with the case of $\bar I =0$.
This automatically ``calibrates" the number of 
injected electrons, and we obtain
\begin{equation}
\bar I_{\rm steady}
=
2 \frac{\hbar k_{\rm F}}{m}
\left[
\frac{\hbar k_{\rm F}}{m^*}
+
\frac{1}{2 \pi} (f_{++} - f_{+-})
\right]^{-1}
\frac{e}{2 \pi \hbar} \Delta \mu
=
\frac{e}{\pi \hbar} \Delta \mu,
\end{equation}
where we have used Eq.\ (\ref{mass}).
Therefore,
$
G = 
e^2/\pi \hbar
$.
Here, the interaction parameters of the 1d field are canceled in $G$, 
and those of the reservoir field are absorbed in $\gamma$.
These observations confirm the results of section \ref{sec_nmf}:
the shifted Fermi state is realized as the nonequilibrium steady state, 
and the conductance is quantized.

The application of the projection theory to 
the case where $\hpn_1$ behaves as a TLL
will be a subject of future study.

\subsection{Advantages of the projection theory}
\label{sec_adv}

A disadvantage of the projection theory is that 
calculations of $G$ become rather hard as compared with 
the simple theories that are reviewed in section \ref{sec_review}.
However, the simple theories have many problems 
and limitations, as discussed there.
The projection theory is free from such problems 
and limitations, and has the following advantages:
(i) The value of $\langle I \rangle$ for the nonequilibrium state 
is directly calculated as a function of $\Delta \mu$.
Hence, neither the translation of
$\Delta \phi_{\rm ext}$ into $\Delta \mu$ 
nor the subtle limiting procedures of $\omega, q$ and $V$ 
is necessary.
(ii) There is no need for the mixing property of 
the 1d Hamiltonian $\hat H_1$. Hence, 
$\hat H_1$ can be the Hamiltonian of 
integrable 1d systems such as the TLL.
(iii) In contrast to the Kubo formula, 
which evaluates transport coefficients from equilibrium fluctuations,
the projection theory gives the nonequilibrium steady state.
This allows us to discuss what 1d state is realized and 
how the current is injected from the reservoirs.
Moreover, we can calculate the NEN and nonlinear responses.
(iv) The projection theory can describe the 
relaxation to the nonequilibrium steady state.
This allows us to study the stability and the relaxation time
of the nonequilibrium state.
%

\section{Appearance of a non-mechanical force}
\label{sec_nmf}

We here discuss the applicability of the Kubo formula
to inhomogeneous systems.
The general conclusion of this section is 
independent of natures (such as a FL or TLL) 
and the dimensionality of the electron system.
Hence, we will use the results for the 1d FL, 
which are obtained in section \ref{sec_FL} and confirmed
by a full statistical-mechanical theory 
in section \ref{sec_proj_ee}.

The original form of the Kubo formula gives 
a conductivity that corresponds to the following conductance;
$
\langle I \rangle / \Delta \phi_{\rm ext}
\equiv 
G_{\rm Kubo}
$ \cite{zubarev,kubo}.
Izuyama suggested that 
the conductance should be
$
\langle I \rangle / \Delta \phi
\equiv 
G_{\rm Izuyama}
$, by considering the screening of $\phi_{\rm ext}$ 
\cite{kawabata,izuyama}.
On the other hand, 
the exact definition of the conductance is 
$G \equiv \langle I \rangle / \Delta \mu$ \cite{callen,zubarev}.
For macroscopic inhomogeneous conductors, 
$e \Delta \phi \neq \Delta \mu$ in general
if one takes the differences between 
both ends of the conductor,
as sketched in Fig.\ \ref{fig_inhomo}(a).
Therefore,  
$G \neq G_{\rm Kubo}, G_{\rm Izuyama}$
in such a case.
Hence, to obtain the correct value of $G$ by the Kubo formula,
one must find the relation between 
$\Delta \phi_{\rm ext}$ and $\Delta \mu$.
Unfortunately, no systematic way of doing this has been developed.

The same can be said for mesoscopic conductors, 
Fig.\ \ref{fig_inhomo}(b), 
for which 
$e \Delta \phi \neq \Delta \mu$ in general
if one takes the differences between 
both ends of the QWR.
Therefore,  
$G \neq G_{\rm Kubo}, G_{\rm Izuyama}$.
It is only for fortunate cases that 
$G_{\rm Kubo}$ or $G_{\rm Izuyama}$ coincides with $G$.
For example, 
Kawabata \cite{kawabata_unpublished} 
calculated $G_{\rm Izuyama}$ for the case
where the backward scattering with amplitude $V(2 k_{\rm F})$ 
is present, which had been neglected in the previous calculations. 
He found that
\begin{equation}
G_{\rm Izuyama}
=
\frac{e^2}{\pi \hbar}
\left[
1 + \frac{V(2 k_{\rm F})}{2 \pi \hbar v_{\rm F}}
\right].
\end{equation}
However, this result disagrees with $G$ obtained
in the previous sections.
The origin of this discrepancy may be understood as follows.
By taking the Fourier transforms of 
both sides of Eq.\ (\ref{ren_phi}),
we can see that only the $q \simeq 0$ component
of the two-body potential $v$ contributes to the screening
of the electrostatic potential. 
On the other hand, both 
$q \simeq 0$ (forward) and $q \simeq 2 k_{\rm F}$ (backward) components
of $v$ contribute the Landau parameter $f_{+-}$, i.e., 
$f_{+-} = f_{+-}^{\rm forward} + f_{+-}^{\rm backward}$.
Therefore, Eq.\ (\ref{DmuFL}) shows that 
$\Delta \mu$ has a term (proportional to $f_{+-}^{\rm backward}$)
which cannot be interpreted as coming from the screening of 
 $\phi_{\rm ext}$.
If we interpret this term in terms of nonequilibrium thermodynamics
(although it is not fully applicable because 
the local equilibrium is not established), 
the term may be interpreted as 
a {\it non-mechanical force} in Eq.\ (\ref{muandmuc}):
\begin{equation}
\Delta \mu_{\rm c} \to - (\hbar q / \pi) f_{+-}^{\rm backward}
+ \cdots.
\end{equation}
Here, $\cdots$ accounts for possible contributions 
from $f_{++}$ and/or $f_{+-}^{\rm forward}$.
Since $q \propto \langle I \rangle$, so is 
$\Delta \mu_{\rm c}$.
This means that 
a finite current $\langle I \rangle$ 
induces a finite non-mechanical force 
$\Delta \mu_{\rm c}$, 
and $\langle I \rangle$ is driven by {\it both} 
$e \Delta \phi$ and $\Delta \mu_{\rm c}$ in the steady state.
Hence,  
$
G 
\equiv 
\langle I \rangle / (\Delta \mu/e) 
$
is {\it not} equal to 
either 
$
G_{\rm Kubo}
\equiv
\langle I \rangle / \Delta \phi_{\rm ext}
$
or
$
G_{\rm Izuyama}
\equiv
\langle I \rangle / \Delta \phi
$.
Therefore,  
the Kubo formula cannot give the correct value
of $G$ if the $q \simeq 2 k_{\rm F}$ component
of the two-body potential is non-negligible,
even if the screening of $\Delta \phi_{\rm ext}$ 
is correctly taken into account,
because a non-mechanical force is inevitably induced.
Note that this is not the unique problem of the Kubo formula, 
but {\it a common problem of many microscopic theories} 
which calculate a nonequilibrium state by applying 
a mechanical force.

A possible way of getting the correct result by 
the Kubo formula would be to apply the formula to 
a larger system that includes the homogeneous reservoirs
or leads \cite{kawabata_great}:
in that case, $e \Delta \phi = \Delta \mu$, as sketched in
Fig.\ \ref{fig_inhomo},
and thus
$G = G_{\rm Izuyama}$.
However, this seems very difficult because 
it is almost equivalent to 
trying to solve the Schr\"odinger equation of the total system, 
including complicated processes that lead to the mixing 
property and to the equality $e \Delta \phi = \Delta \mu$.
Another possible solution may be to apply Zubarev's 
method \cite{zubarev},
which, 
to the authors' knowledge, 
has not been applied to interacting electrons 
in mesoscopic conductors.
However, one must also include (a part of) reservoirs into
the Hamiltonian because Zubarev's method assumes 
that macroscopic variables (such as $\mu$) are well-defined
in the nonequilibrium steady state. 
As compared with these approaches, 
the formulations presented in sections 
\ref{sec_combined} and \ref{sec_proj}
would be simpler ways of getting correct results 
which include effects of non-mechanical forces.

\section{Deviation from the quantized conductance}
\label{sec_deviation}

For a clean QWR, 
we have obtained the quantized value $G = e^2/\pi \hbar$ 
in both cases of the FL and the TLL in sections
\ref{sec_FL}, \ref{sec_TLL} and \ref{sec_proj_ee},
using different formulations.
The essential assumptions leading to this result are the following.
(i) The QWR is clean enough and the temperature is low
enough (zero temperature has been assumed for simplicity),
so that scatterings by impurities, defects or phonons
are negligible and e-e interactions 
are the only scattering mechanism.
(ii) The boundaries between
the QWR and reservoirs
are smooth and slowly-varying 
so that reflections at the boundaries are absent.
(iii) The reservoirs are large enough,
so that they remain at equilibrium even in the presence of
a finite current between the reservoirs through the QWR.

When some of these assumptions are not satisfied
the observed conductance may deviate from the quantized value.
For example, if boundary reflections are non-negligible,
the transmittance $T$ (calculated from the single-body Schr\"odinger 
equation) between the reservoirs through the QWR
is reduced.
This results in the reduction of 
$G$ by the factor $T$ for non-interacting electrons.
For interacting electrons, $G$ will be further reduced
for the TLL because the TLL will be ``pinned" 
by the reflection
potential at the boundaries.
This can be understood simply as follows:
Although the TLL of infinite length is a liquid, for 
which a long-range order is absent,
it behaves like a solid at a short distance.
Hence, the TLL is pinned by a local potential, like a charge-density 
wave is.
This is the physical origin of the vanishing $G$ (at zero temperature)
for the case where a potential 
barrier is located {\it in} the TLL \cite{KF,FN}.
Since the pinning occurs irrespective of the position of the 
local potential, the TTL would be pinned also by 
the boundary reflections.
Note that 
if one neglects the weakening of $v_{11}$ 
in reservoirs (due to the broadening of $W(x)$,
as shown in section \ref{sec_H}),
the TTL would then be pinned also by 
impurities in reservoirs \cite{kawabata_great}.

Another example is dissipation by, say, phonon emission.
By the dissipation, 
the 1d system will lose any correlations over a distance $L_{\rm rlx}$,
where $L_{\rm rlx}$ is the
``maximal energy relaxation length" \cite{su92,isqm92},
which is generally longer than the simple dephasing length (over which
an energy correlation may be able to survive).
In such a case the 1d system of length $L$ ($> L_{\rm rlx}$) will behave as
a series of independent conductors of length $L_{\rm rlx}$.
One will then observe Ohm's law \cite{s96}:
\begin{equation}
G_{\rm obs} \simeq (L_{\rm rlx}/L)\times(e^2/\pi \hbar).
\end{equation}

%

\subsubsection{Acknowledgment}
The authors are grateful to helpful discussions with
A. Kawabata, T. Arimitsu, and K. Kitahara.
This work has been supported by the Core Research for 
Evolutional Science and Technology (CREST) of the Japan Science 
and Technology Corporation (JST).

%


\begin{thebibliography}{8.}
\bibitem{why_e}
The role of the factor $1/e$ is just to convert 
the unit of $\Delta \mu$ into volt.
\bibitem{nontrivial}
It is sometimes argued that for a system with the translational
symmetry the $e$-$e$ interaction would not affect the value of $G$.
For a mesoscopic conductor, however, $\Delta \mu$ (hence $G$) can be 
defined only when 2d or 3d reservoirs are connected, which 
violates the translational symmetry. 
\bibitem{def_of_mu}
In thermodynamics, $\mu$ is called an 
``electrochemical potential'', whereas
$\mu_{\rm c}$ is called a chemical potential \cite{callen}.
We here follow the terminology used in the solid state physics,
where $\mu$ is called a chemical potential
\cite{AR,KF,FN,OF,su92,maslov,pono,kawabata,safi,oreg96,LB,s96}.
%
Note that it is not $\Delta \phi$ but $\Delta \mu$ that 
is applied by a battery.
\bibitem{exception}
The only exception is the NEN of (tunnel) junctions with a high barrier.
In this case one can take the approximate equilibrium state with 
$\Delta \mu > 0$ as the equilibrium state of the Kubo formula.
\bibitem{note_mu}
Note that this equality should be satisfied only under the 
condition (which we assume) that the boundary reflections
are negligible.
\bibitem{adiabatic}
This is an adiabatic approximation, which has been 
widely used in studies of optical waveguides.
The equation for the optical field in a waveguide has 
the same form as the single-body Schr\"odinger equation 
of an electron in a quantum wire.
\bibitem{2nd}
Note that the second-order treatment of $\hat V_{\rm 1R}$ 
does not limit the accuracy of the theory
because, e.g., the result for the steady current
does not include $\hat V_{\rm 1R}$.
Only the accuracy of the relaxation time $\gamma$ is 
limited to the second order.
\bibitem{precise}
Precisely speaking,
the equality
$\hat \zeta_1(t') \simeq \hat \zeta_1(t)$ 
holds not for $\hat \zeta_1$ but 
for expectation values evaluated from $\hat \zeta_1$.
We therefore apply Eq.\ (\ref{master}) to expectation values, 
as demonstrated in sections 
\ref{sec_proj_impurity} and \ref{sec_proj_ee}.
\end{thebibliography}
\end{document}